\documentclass[aps,prd,amsfonts,amsmath,nofootinbib,showpacs,preprint,tightenlines,longbibliography]{revtex4-1} 

\def\be{\begin{equation}}
\def\ee{\end{equation}}
\def\bea{\begin{eqnarray}}
\def\eea{\end{eqnarray}}
\def\bml{\begin{subequations}}

\def\elea{\end{eqnarray}\end{subequations}}

\def\bX{\mathbf{X}}
\def\bP{\mathbf{P}}
\def\Xd{\dot\bX}
\def\Xdd{\ddot\bX}
\def\Xddd{\dddot\bX}
\def\nn{\mathbf{\hat n}}

\def\JS{Ref.~\cite{Jenkins:2020ctp}}

\begin{document}

\title{No black holes from cosmic string cusps}

\author{Jose J. Blanco-Pillado}
\email{josejuan.blanco@ehu.es}
\affiliation{Department of Physics, UPV/EHU,\\48080, Bilbao, Spain}
\affiliation{IKERBASQUE, Basque Foundation for Science, 48011, Bilbao, Spain}

\author{Ken D. Olum}
\email{kdo@cosmos.phy.tufts.edu}

\author{Alexander Vilenkin}
\email{vilenkin@cosmos.phy.tufts.edu}
\affiliation{Institute of Cosmology, Department of Physics and Astronomy,\\Tufts University, Medford, MA 02155, USA}

\begin{abstract}
Recent work by Jenkins and Sakellariadou \cite{Jenkins:2020ctp} claims
that cusps on cosmic strings lead to black hole production.  To derive
this conclusion they use the hoop conjecture in the rest frame of the
string loop, rather than in the rest frame of the proposed black hole.
Most of the energy they include is the bulk motion of the string near
the cusp.  We redo the analysis taking this into account and find that
cusps on cosmic strings with realistic energy scale do not produce
black holes, unless the cusp parameters are extremely fine-tuned.
\end{abstract}

\maketitle

Suppose there is a cosmic string network with tension and linear
energy density $\mu$.  Is it possible that black holes can form from
concentrations of energy in this string network?  The gravitational
effects of strings depend on the dimensionless parameter $G\mu$, where
$G$ is Newton's constant, and we work in units where the speed of
light $c = 1$.  All realistic models have $G\mu \lesssim 10^{-7}$ to
avoid conflict with cosmic microwave background observations
\cite{Ade:2013xla}, and in the usual case of gauge strings that decay
only by gravitational wave emission the bound is much stronger
(e.g., see Refs.~\cite{Sanidas:2012ee,Blanco-Pillado:2017rnf,Abbott:2017mem}.)
Thus a typical string loop is very far from becoming a black hole.  It
requires an enhancement of order $1/(G\mu)$ in the energy density to
raise the possibility of gravitational collapse.

Collapse to a black hole is certainly possible in special cases, for
example a momentarily static, circular string
\cite{Helfer:2018qgv,Aurrekoetxea:2020tuw}.  However, the string must
be very accurately circular, and we expect such alignment to be
extraordinarily rare
\cite{Hawking:1987bn,Polnarev:1988dh,Caldwell:1993kv,Caldwell:1995fu,Hansen:1999su,James-Turner:2019ssu}.
But Jenkins and Sakellariadou~\cite{Jenkins:2020ctp} suggest instead
that cusps, which are a generic feature of cosmic string loops where a
point on the string momentarily moves (formally) at the speed of light
\cite{Turok:1984cn}, will frequently lead to black hole formation.
Their analysis is based on Thorne's ``hoop conjecture'', which states
that ``Horizons form when and only when a mass $M$ gets compacted into
a region whose circumference in every direction is $\mathcal{C} < 4\pi
GM/c^2\,$'' \cite{Thorne:1972ji-special}.  The original conjecture was
intended to apply to curved spacetime, but \JS{} did not consider
curvature arising before the time at which the above condition was
satisfied, and we will do the same here.  In that case it becomes
simply a statement about the mass $M$ inside a ball of radius $r$,
saying that a black hole will form if $r$ and $M$ satisfy the ``hoop
condition''~\cite{Jenkins:2020ctp},
\be
\frac{2 G M}{r} > 1\,.
\ee

The hoop conjecture has never been rigorously formulated, but it is
obvious that whether or not a black hole forms cannot depend on the
inertial frame from which the process is viewed.  It is thus incorrect
to include the center-of-mass kinetic energy of a moving object in the
mass-energy that might satisfy the conjecture.  If one includes it,
one would conclude that any object would collapse into a black hole,
because it can always be viewed from a highly boosted Lorentz frame in
which it will have an arbitrarily large total energy.  Instead one
should apply the conjecture in the rest frame of the material whose
possible collapse one is considering.

In the case of a cusp, there is no ``rest frame of the cusp'', because
the tip moves at the speed of light in any frame.  But if we consider
a certain segment of the string near the cusp that might form a black
hole, we should test the hoop conjecture in the rest frame of that
segment.

We will describe the string in terms of left-moving and right-moving
excitations as usual, with the same notation as \JS,
\be
\bX(t,\sigma) = {\frac12} \left[\bX_-  (\sigma_-) + \bX_+ (\sigma_+)\right]\,,
\ee
where $\sigma$ parameterizes ``invariant length'' (energy in units of
$\mu$), $\sigma_\pm = t\pm \sigma$, and the functions
$\bX_\pm(\sigma)$ have unit tangent vector.  The string velocity is
\be
\dot\bX(t,\sigma) = {\frac12} \left[\Xd_-  (\sigma_-) + \Xd_+ (\sigma_+)\right]\,,
\ee
where dots mean derivatives of $\bX$ with respect to time and
$\bX_\pm$ with respect to its argument $\sigma_\pm$.

We take the cusp to be at $t = \sigma =0$ and
expand the string around the
cusp to third order,
\be\label{eqn:Taylor}
\bX_\pm (\sigma_\pm) = \nn\sigma_\pm + \frac12 \Xdd_\pm \sigma^2_\pm + \frac16 \Xddd_\pm \sigma^3_\pm\,,
\ee
where $\nn$ is a unit vector pointing in the direction that the cusp
tip moves, and the derivatives are evaluated at the cusp.  To maintain
the unit tangent vector, $|\Xd_\pm(\sigma)|= 1$, requires $\nn\cdot\Xdd_{\pm} =0$
and $\nn\cdot\Xddd_{\pm} =-|\Xdd_\pm|^2$.  Following \JS{} we consider a
snapshot at $t = 0$ and ask whether the region of string with
$-\sigma_* < \sigma < \sigma_*$ satisfies the hoop condition.  We will
let $r$ be the maximum distance of the string from the point of the
cusp.  This is dominated by the second-order term,
\be\label{eqn:r}
r = \frac12 |\Xdd|\sigma_*^2\,.
\ee
The energy in the range $-\sigma_*\ldots\sigma_*$
(called $M$ in \JS) is just $E = 2\mu\sigma_*$.  
As we decrease $\sigma_*$, $E$ grows relative to $r$.  If
we use $E$ in the hoop condition, we satisfy it at
\cite{Jenkins:2020ctp}
\be
\sigma_*^{\text{BH}} = \frac{8G\mu}{|\Xdd|}\,,\qquad
r^{\text{BH}} = \frac{32G^2\mu^2}{|\Xdd|}\,,\qquad
E^{\text{BH}} = \frac{16G\mu^2}{|\Xdd|}\,.
\ee

A typical value of $|\Xdd|$ would be around $1/\ell$, where $\ell$ is
the invariant length of the loop.  Thus $\sigma_*$ is of order 
$G\mu\ell$, so by this criterion a black hole would form containing a
fraction about $G\mu$ of the original loop \cite{Jenkins:2020ctp}.

But most of the energy appearing in $E$ above is the kinetic energy of
the center-of-mass motion of the string near the cusp.  This should
not be included in the hoop condition.  Since the string moves
primarily in the direction of $\nn$, let us calculate the total momentum
in that direction.  The $\nn$ velocity of the string is
\be
\nn\cdot\Xd = 1 + \frac12\nn\cdot\Xddd~\sigma^2
= 1 - \frac{1}{4}(|\Xdd_+|^2 + |\Xdd_-|^2) \sigma^2\,,
\ee
and the total $\nn$ momentum is
\be\label{eqn:momentum}
\nn\cdot\bP = \mu\int_{-\sigma_*}^{\sigma_*} \nn\cdot\Xd \, d\sigma
= 2\mu\sigma_* - \frac{\mu}{6}(|\Xdd_+|^2 + |\Xdd_-|^2)\sigma_*^3\,.
\ee
The total momentum $P = |\bP|$ is at least $\nn\cdot\bP$, so the rest
mass is limited by\footnote{In fact \JS{}
  computed the rest mass of the black hole but continued to use the
  total energy as the mass in the hoop conjecture.}
\be
M = \sqrt{E^2-P^2} < \frac{2\mu\sigma_*^2}{\sqrt{6}}
\sqrt{|\Xdd_+|^2 + |\Xdd_-|^2}
= \frac{4\mu r}{\sqrt{6}}
\frac{\sqrt{|\Xdd_+|^2 + |\Xdd_-|^2}}{|\Xdd|}\,.
\ee
Now $M$ depends linearly on $r$, and satisfying the hoop condition
using the rest mass would require
\be\label{eqn:condition}
\frac{|\Xdd|}{\sqrt{|\Xdd_+|^2 + |\Xdd_-|^2}} < \frac{8G\mu}{\sqrt{6}}\,.
\ee
Since $G\mu <10^{-7}$, Eq.~(\ref{eqn:condition}) can never be
satisfied except in a pathological case where $\Xdd$ is tiny.  This
requires the lengths of $\Xdd_+$ and $\Xdd_-$ to be the equal to
accuracy $G\mu$ and also the angle between them to be small
on the order of $G\mu$.  (There is only one angle because both are
perpendicular to $\nn$.). Thus we would expect only a fraction of cusps
of order $(G\mu)^2$ to form black holes, so that the number
of black holes expected from these models is extremely suppressed.

The above analysis is not entirely correct, because we considered a
snapshot of the string in the ``laboratory'' frame, rather than in the
frame where we want to apply the hoop conjecture.  Instead we can view
a segment of the cusp in a frame where it has no overall momentum.  It
is still a cusp and we can write the Taylor series as in
Eq.~(\ref{eqn:Taylor}), where the parameter $\sigma$ now parameterizes
the energy as seen in the moving frame.  The transformation of the
coefficients is considered in detail in
Ref.~\cite{BlancoPillado:1998bv}.  Unfortunately, in this case, the
Taylor series may not be dominated by the first three orders given
in Eq.~(\ref{eqn:Taylor}).  We would like to use
Eq.~(\ref{eqn:Taylor}) out to some new $\sigma_*$ where the total
momentum of Eq.~(\ref{eqn:momentum}) would vanish.  That implies
\be
\sigma_*^2 = \frac {12}{|\Xdd^{CF}_+|^2 + |\Xdd^{CF}_-|^2}\,,
\ee
so $\Xdd^{CF}$ is of order $\sigma_*$.  Then $\Xddd^{CF}$ is of order
$\sigma_*^2$, and there is no small parameter that would suppress
higher-order terms\footnote{Note that these vectors are evaluated in the
"cusp frame" so their magnitudes are not simply set by the overall
size of the loop as in the "laboratory frame".}.

Nevertheless we expect this analysis to give the correct order of
magnitude.  The mass of the string from $-\sigma_*$ to $\sigma_*$ is
just $2\mu\sigma_*$, while the size of the region is at least as large
as the $r$ given by Eq.~(\ref{eqn:r}).  The hoop condition then gives
something analogous to Eq.~(\ref{eqn:condition}), with a different
numerical factor.  There may also be an additional dependence on the
relative magnitudes of $\Xdd^{CF}_+$ and $\Xdd^{CF}_-$, and the angle between
them.  The left-hand side of Eq.~(\ref{eqn:condition}) is invariant
under boosts \cite{BlancoPillado:1998bv}, and is generally of order 1,
so any such bound could only be satisfied by extreme fine-tuning, as
above.\\

Thus we see that there is no significant formation of black holes from
cosmic string cusps.

\section*{Acknowledgements}
We would like to thank Yuri Levin and Andrei Gruzinov for helpful 
conversations. This work is supported in part by the National Science Foundation grants, PHY-1820872 and  
PHY-1820902, the Spanish Ministry MINECO, MCIU/AEI/FEDER grant (PGC2018-094626-B-C21),
the Basque Government grant (IT-979-16) and the Basque Foundation for Science (IKERBASQUE).

\bibliography{paper,no-slac}

\end{document}